\documentclass[manuscript]{aastex}

\slugcomment{To appear in the Astrophysical Journal Letters} 
\shorttitle{Var C in M33 }
\shortauthors{Humphreys et al. }

\begin{document}

\title{The Wind of Variable C in M33\altaffilmark{1} }

\author{
Roberta M.. Humphreys\altaffilmark{2}, 
Kris Davidson\altaffilmark{2}, 
Michael Gordon\altaffilmark{2},  
Kerstin Weis\altaffilmark{3},
Birgitta Burggraf\altaffilmark{3},  
D.~J. Bomans\altaffilmark{4},
and
John C. Martin\altaffilmark{5} 
}

\altaffiltext{1}  
{Based  on observations  with the Multiple Mirror Telescope, a joint facility of
the Smithsonian Institution and the University of Arizona and on observations obtained with the Large Binocular Telescope (LBT), an international collaboration among
 institutions in the United
 States, Italy and Germany. LBT Corporation partners are: The University of
 Arizona on behalf of the Arizona university system; Istituto Nazionale di
 Astrofisica, Italy; LBT Beteiligungsgesellschaft, Germany, representing the
 Max-Planck Society, the Astrophysical Institute Potsdam, and Heidelberg
 University; The Ohio State University, and The Research Corporation, on
 behalf of The University of Notre Dame, University of Minnesota and
 University of Virginia.
 .
} 

\altaffiltext{2}
{Minnesota Institute for Astrophysics, 116 Church St SE, University of Minnesota
, Minneapolis, MN 55455; roberta@umn.edu} 

\altaffiltext{3}
{Astronomical Institute, Ruhr-Universitaet Bochum, Germany, }

\altaffiltext{4}
{Astronomical Institute, Ruhr-Universitaet Bochum, Germany and RUB Research Department "Plasmas with complex interactions",}

\altaffiltext{5}
{Barber Observatory, University of Illinois, Springfield, IL, 62703}

\begin{abstract}
We discuss the spectrum of Var C in M33 obtained just before the onset of its current 
brightening and  recent spectra during its present ``eruption'' or optically thick wind stage. These 
spectra illustrate the typical LBV transition in apparent spectral type or temperature
that characterizes the classical LBV or S Dor-type variability. LBVs are known to have 
slow, dense winds during their maximum phase. Interestingly, {\it Var C  had a slow wind even during its hot, quiescent stage in comparison with the normal hot supergiants with 
similar temperatures. Its outflow or wind speeds also show very little change 
between these two states.} 
\end{abstract} 


\section{Introduction}

Variable C in M33, one of the original Hubble-Sandage variables is a known S Doradus
variable/classical Luminous Blue Variable (LBV) in M33 \citep{HS,RMH88,Szeif}. 
A  classical LBV in quiescence (minimum  visual light) resembles a moderately evolved hot star with the spectrum of an early B-type supergiant or Of/late WN star \citep{HD94}. During an LBV eruption, the mass loss rate increases, the wind becomes opaque and cool, and its spectrum
resembles an A -- F-type supergiant. Since this is a shift in the bolometric correction, the
star brightens in the visual and appears to move to the right, to lower temperatures on the
HR Diagram. This is the LBV's optically thick wind stage or maximum  visual light.

Var C's historic light curve is  discussed  by \citet{Burg}. It has shown two relatively long periods of 
maximum  light; from  1940 - 1953 \citep{HS} and 1982 - 1993 \citep{RMH88,Szeif} with two shorter maxima in  1964 - 1970 \citep{Ros} and again beginning in 2001 and lasting until about  2005 based on the 
photometry and spectra reported by \citet{Viotti}.   
Var C entered another LBV eruption or maximum light phase apparently beginning  
close to  2011.0  when it  brightened about two magnitudes in the visual 
reaching V $=$ 15.6 mag \citep{RMH_VarC2013}. 

In this {\it Letter} we describe its spectrum  obtained just before 
the onset of its current brightening and recent spectra from its present maximum. 
Interestingly, we find that its outflow or wind speed shows little change between these two 
states and is slow even during quiescence; much slower than the winds of comparable B-type 
supergiants. This observation may be a significant clue to the LBV instability in general.  

\section{Observations} 

Variable C was observed on 03 October 2010 and on 07 October 2013 with the Hectospec Multi-Object Spectrograph (MOS) \citep{Fab98} on the 6.5-m MMT on Mt. Hopkins as part of a larger program 
on luminous stars in M31 and M33 \citep{RMH13,RMH14}. The Hectospec\footnote{http://www.cfa.harvard.edu/mmti/hectospec.html} has a 1$\arcdeg$ FOV and uses 300 fibers each with a core 
diameter of 250$\mu$m subtending 1$\farcs$5 on the sky. We used the 600 l/mm grating with the 
4800{\AA} tilt  yielding  $\approx$ 2500{\AA} coverage with 0.54 {\AA}/pixel resolution 
and R of $\sim$  2000. The same grating with a tilt of 6800{\AA} was used for the red 
spectra with $\approx$ 2500{\AA} coverage, 0.54{\AA}/pixel resolution and R of $\sim$ 3600.  
The spectra  were reduced using an exportable version of the CfA/SAO SPECROAD package 
for  Hectospec data\footnote{External SPECROAD was developed by Juan Cabanela for use
on Linux or MacOS X systems outside of CfA. It is available online at:
http://iparrizar.mnstate.edu.}. The spectra were all bias subtracted, flatfielded and 
 wavelength calibrated. Because of crowding, the sky subtraction was done using 
 fibers assigned outside the field of the galaxy.
Blue and red spectra were also observed on 06 January 2014 with the MODS1 
spectrograph on the Large Binocular Telescope (LBT). MODS1 uses a dichroic to 
obtain blue and red spectra simultaneously with the  G400L and G750L gratings   
for the blue and red channels, respectively, yielding  wavelength coverage from  3200{\AA} in the blue to more than 1$\mu$m  in the red. The spectra were reduced using a pipleine provided by R. W. Pogge for MODS spectra and standard routines in IRAF.

Figure 1 shows the recent light curve based on photometry from  \citet{Burg}, \citet{Viotti13},  \citet{Valeev}, and from  Martin's CCD observations with the 20-inch telescope at the Barber Observatory.  
Although there are gaps in the present photometry, we suspect that the current 
eruption  began approximately 2011.0 rather than in mid-2010. This is on based Var C's  spectrum  in  Oct 2010 when it was still in its hot or quiescent state, and because at
the beginning of the 1982 -- 1993 eruption it showed a rapid rise to maximum in only a few weeks \citep{RMH88}. 

Soon after the announcement that Var C had brightened, several groups obtained low 
resolution spectra of Variable C. All of the known spectroscopic observations beginning 
with the  2010 spectra are listed in
Table 1. Both sets of our blue and red spectra from the MMT/Hectospec 
are shown in Figure 2. 

\section{The Spectrum of Var C}

Var C's spectrum  in October 2010 shows the absorption line spectrum of an early B-type supergiant plus emission  lines of hydrogen with P Cygni profiles, and He I, Fe II and [Fe II]. The multiplet 42 Fe II 
lines also have prominent P Cygni absorption profiles. The presence of absorption lines  
of N II 3995{\AA}, several He I absorption lines and S IV 4089{\AA} in the blue suggest a 
corresponding B1 - B2 spectral type. The N II lines 4600 --4640{\AA} and 5666 -- 5710{\AA}  
and S III 4553 -- 4575 {\AA} are also present in absorption. The relative strengths of the 
nebular [N II] emission lines indicate that Var C has a circumstellar nebula \citep{Weis13}.

Its spectrum  three years later, after the rise to maximum 
light, shows the transition in apparent spectral type that characterizes the classical 
LBV or S Doradus variability (Figure 2). Its absorption line spectrum  now resembles a late A-type supergiant with
 prominent Ca II H and K absorption lines, strong Mg II $\lambda$4481{\AA}, and  numerous 
 absorption lines of Fe II. The luminosity sensitive O I triplet at $\lambda$7774{\AA}  
 is quite strong. The Si II doublet at 6347 and 6371{\AA} and the three N I lines from 7423 -- 7468{\AA} are present in absorption in its red spectrum. The He I emission and absorption lines are gone. 
The Balmer emission line strengths have decreased, by a factor of two in equivalent width for H$\alpha$. The hydrogen  lines have  developed broad wings extending to $\approx$ 800 km s$^{-1}$, most likely due to Thomson scattering in the dense wind, and the P Cygni absorption is now much stronger indicating that the mass loss rate has increased.
The wings contain roughly 20\% of the H$\alpha$ and H$\beta$ emission (Figure 3).
Likewise the lines of Fe II multiplet 42 have  prominent P Cygni absorption features and 
asymmetric red  wings, a characteristic of Thomson scattering. 
A spectrum  from January 2014  confirms the above description and also  shows the Ca II 
triplet in emission in the far red; two of the Ca II line have P Cygni profiles. 

Thomson-scattered line wings almost automatically appear when 
a hot star's mass outflow becomes opaque.  In such a case the diffuse 
continuum photosphere roughly coincides with the thermalization 
depth, located where    
$3 \, \tau_\mathrm{tot} \, \tau_\mathrm{abs} \; \sim \; 1 \, $.   
With likely opacities, this occurs near scattering depth  
$\tau_\mathrm{sc} \; \sim \;$ 2 to 5 \citep{dav87}.    
Naturally the strongest emission lines are formed in the region outside 
the diffuse photosphere, i.e., with an average $\tau_\mathrm{sc}$ 
of the order of 1 or 2;  and this is about the right depth to produce 
the observed line wings.  \citet{sn2011ht} have 
noted this generality for other types of objects.

Var C was observed during its previous extended maximum, 1982 -- 1993  \citep{RMH88}. Comparison with  spectra from  that time shows  that the current spectrum is  somewhat warmer than the
1985 spectrum  obtained when the star was  about two  years past the onset of the 
eruption. At that time, it 
 had the spectral characteristics and absorption line strengths  of an early F-type 
supergiant.  Instead the current spectrum  is more like  spectra from 1986 and 
1987 when it
 resembled an A-type supergiant and had an apparent temperature of $\approx$ 9000K. 
This is consistent with the light curve.  Like other well-observed LBVs in eruption, Var C's apparent spectrum/temperature is correlated with the amplitude of its visual brightening \citep{Stahl}. Var C has not gotten as visually bright as it did 
in 1985. As of this writing, its maximum visual magnitude is 15.6 compared to 15.2 mag in 1985. Its brightness and colors are essentally the same as the 1986/1987 photometry and the spectra  are likewise very similar. Thus the dense wind during this   maximum  may not get as cool, although of course that will depend on how the 
light curve develops in the future.

\section{The Wind and Outflow Velocity}

LBV's are well known to have low wind or outflow velocities of 100 - 200 km s$^{-1}$ in their cool, dense winds during their eruptions or maximum  light phase. To examine 
Var C's wind,  we measured the velocity at the blue-edge of the P Cygni absorption profiles in the Balmer and Fe II emission lines. This velocity is usually referred to as the terminal 
velocity ($v_{\infty}$), but since we are not using a stellar wind model to fit the profiles, here  we call it the blue-edge velocity. We also measured the velocity at the absorption minimum  
in the P Cygni profiles. This second method is less subjective than the blue-edge  and permits a well-controlled differential measurement. The velocity of the absorption minimum  is of course smaller, but  when comparing the velocities from different times, the results are consistent with the behavior of the terminal or blue-edge velocity. The average velocities with their standard deviations are
summarized in Table 2 for the quiescent and maximum light spectra.   
The H$\alpha$ and H$\beta$ line profiles at the two epochs are shown in Figure 3. 

 For comparison, 
we also include the wind speeds from  the P Cygni profiles in the Balmer lines in the spectra of normal early B-type and late A-type supergiants in M31 and M33 observed with the Hectospec at the same time as Var C in 2010 (\S {1} and Paper I). The line profiles were measured the same way as for Var C.  Wind speeds are expected to have some dependence on metallicity. Although we do not have a measured metallicity for Var C, we expect it to be similar to the other supergiants in M33 which has a metallicity like the LMC while M31 stars have Galactic  abundances .   
Furthermore we do not find a significant difference in the wind velocities for the M31 and M33 stars in this small sample. 

Var C's outflow velocities are significantly less, by at least 70 km s$^{-1}$ or about 30\%  than for the normal B-type supergiants in M31 and M33 even in quiescence when the LBV's apparent spectral type and spectroscopic temperature are like 
that of the hot supergiants\footnote{The wind speeds of Galactic 
and LMC early-type supergiants range from 400 to 1000 km  s$^{-1}$ \citep{Crowther,Mokiem}}.  During the dense wind or maximum  light phase the outflow velocities are 
similar to those of the A-type supergiants, which Var C resembles spectroscopically, although still somewhat lower. {\it Var C thus has a slow wind even during the quiescent stage in comparison with the normal hot supergiants with similar temperatures.
There is also only a small decrease in the wind speeds between these two states.}  

We also find a relatively low wind velocity in the other LBVs in M31 and M33 in their hot or quiescent state, discussed in Paper II in our series on the luminous
stars in these galaxies \citep{RMH14}. Thus  a slower and presumably denser wind, even in quiescence or minimum light, may be another distinguishing characteristic of the classical LBVs/S Doradus variables.  
 A hot star should have a wind speed related  to the 
escape velocity. LBVs are presumably close to the Eddington limit \citep{HD94} having already
shed much of their mass in previous S Dor-type maxima or possibly in enhanced mass loss during a previous giant 
eruption. Their effective gravities  and therefore their escape velocities 
are now much lower.  
The Var C observations support this concept, i.e. that it is fairly close to the Eddington limit. 

We thank U. Munari and R. Mark Wagner for communicating the results of their early 
spectra with us.
Research by R. Humphreys and K. Davidson on massive stars is supported by  
the National Science Foundation AST-1019394. J. C . Martin's collaborative work on luminous variables is supported by the National Science Foundation grant  AST-1108890.

{\it Facilities:} \facility{MMT/Hectospec, LBT/MODS1}


\begin{deluxetable}{lllll}
\tablewidth{0 pt}
\tabletypesize{\footnotesize}
\tablenum{1} 
\tablecaption{Spectroscopic Observations of Var C 2010 -- 2013}
\tablehead{
\colhead{UT Date} &
\colhead{Spectrograph}  &
\colhead{Exp. Time} &
\colhead{Grating/Tilt or  Wavelength} &
\colhead{Reference} 
}
\startdata 
29 Sep --02 Oct, 2010  &  WHT/WYFFOS  &  9H, 6.5H  &  4000--5000{\AA}, H$\alpha$    & \citet{Clark12}\\
03 Oct. 2010  & MMT/Hectospec & 120m  &  600l, 4800{\AA} &  This paper \\
03 Oct. 2010  & MMT/Hectospec &  90m  &  600l, 6800{\AA} & This paper \\
05 Sept. 2013 & MDM/CCD       &  30m  &  3650 -- 7250{\AA} & \citet{Wagner}\\ 
06 Sept. 2013 & Asiago 1.2m /B\&C   &  30m  &  300l, 3400--7980{\AA} &  \citet{Munari}
 \\
 05 Oct. 2013  & BTA/SCORPIO   &  \nodata &  3800 -- 7200{\AA} & \citet{Valeev} \\
 07 Oct. 2013  & MMT/Hectospec & 120m  &  600l, 4800{\AA} &  This paper \\
 07 Oct. 2013  & MMT/Hectospec &  90m  &  600l, 6800{\AA} & This paper \\ 
 01 - 02 Nov. 2013 &   BTA/SCORPIO & \nodata   &  4100 -- 5800, 5800 -- 7400{\AA} & \citet{Valeev} \\
 07 Nov. 2013 & Asiago 1.2m /B\&C   &  30m  &  300l, 3400--7980{\AA} &  \citet{Munari} \\  
 06 Jan. 2014 & LBT/MODS1  &  6m    & 3200{\AA} -- 1$\mu$m  &  This paper \\ 
\enddata
\end{deluxetable}

\begin{deluxetable}{lcccc}
\tablewidth{0 pt}
\tabletypesize{\footnotesize}
\tablenum{2} 
\tablecaption{Var C Outflow Velocities (km s$^{-1}$) \tablenotemark{1} }
\tablehead{
\colhead{Star} &
\colhead{P Cyg (H)} &
\colhead{P Cyg (Fe II)} & 
\colhead{P Cyg (H)} &
\colhead{P Cyg (Fe II)}\\
   & 
\colhead{blue-edge}&
\colhead{blue-edge}&
\colhead{abs. min.}&
\colhead{abs. min.} 
}
\startdata
Var C Oct. 2010 & $-$238(4)$\pm$ 10 &  $-$249(3)$\pm$ 15 & $-$157(4)$\pm$ 8 &  $-$158(3)$\pm$ 3 \\
    
B-type supergiants(B0-B3)\tablenotemark{2} & $-$351(7)$\pm$ 9 & \nodata & $-$227(7)$\pm$ 9 & \nodata \\
Var C Oct. 2013\tablenotemark{3} & $-$233(4)$\pm$ 10 &  $-$234(8)$\pm$ 8 & $-$141(4)$\pm$ 3 &  $-$136(8)$\pm$ 3 
\\ 
A-type supergiants(A5-A8)\tablenotemark{4} & $-$266(6)$\pm$ 20 & \nodata  & $-$162(7)$\pm$ 14 & \nodata \\ 
\enddata
\tablenotetext{1}{The number of lines measured is given in parenthesis.}
\tablenotetext{2}{Three stars, 2 M33, 1 M31} 
\tablenotetext{3}{The outflow velocites from  the absorption minimum  in the two Ca II lines is -148 km s$^{-1}$.} 
\tablenotetext{4}{Four stars, 2 M33, 2 M31}
\end{deluxetable}


\begin{figure}
\figurenum{1}
\epsscale{1.0}
\plotone{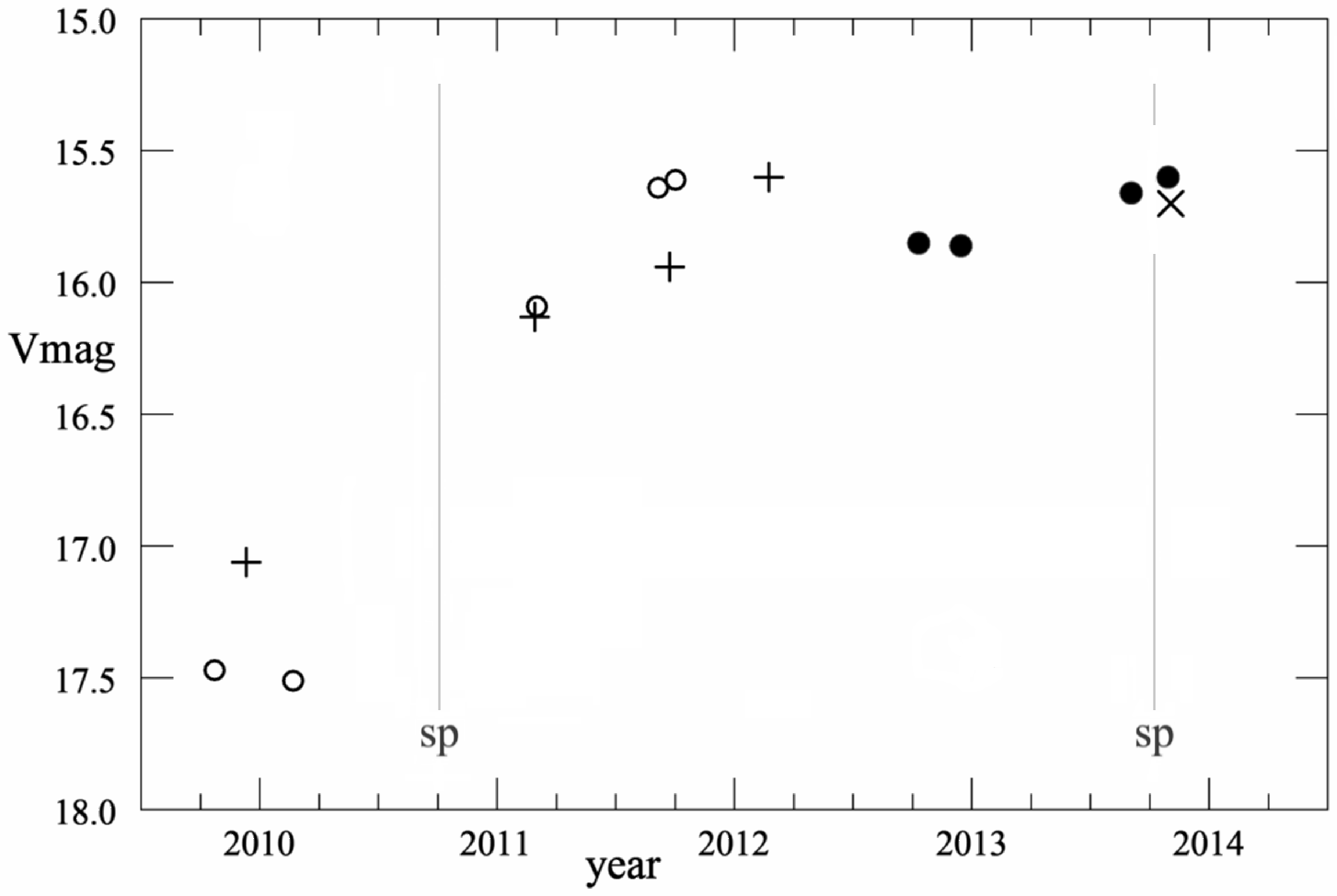}
\caption{The light curve of Var C from late 2009 to the present.The different symbols are from different sources; $\bullet$ this paper, $\circ$ \citet{Burg}, +  \citet{Viotti13}, and $\times$ \citet{Valeev}. The vertical lines mark the dates of the MMT spectra. }
\end{figure}

\begin{figure}
\figurenum{2}
\epsscale{1.0}
\plotone{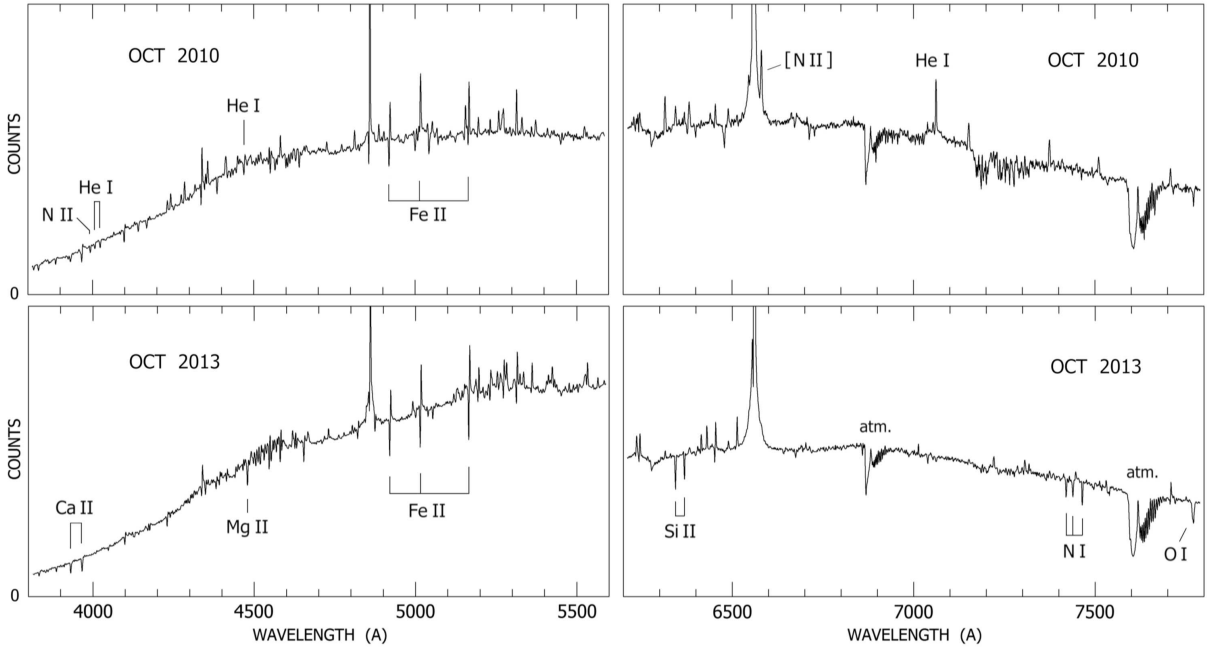}
\caption{The spectra of Var C in 2010 before the onset of the eruption and a recent 2013  spectrum  during the current maximum.. The {\it absorption} lines illustrate the change in apparent spectral type. The complex absorption at 7200 --7300{\AA} in 2010 is H$_{2}$O in the Earth's atmosphere. }
\end{figure}

\begin{figure}
\figurenum{3}
\epsscale{0.6}
\plotone{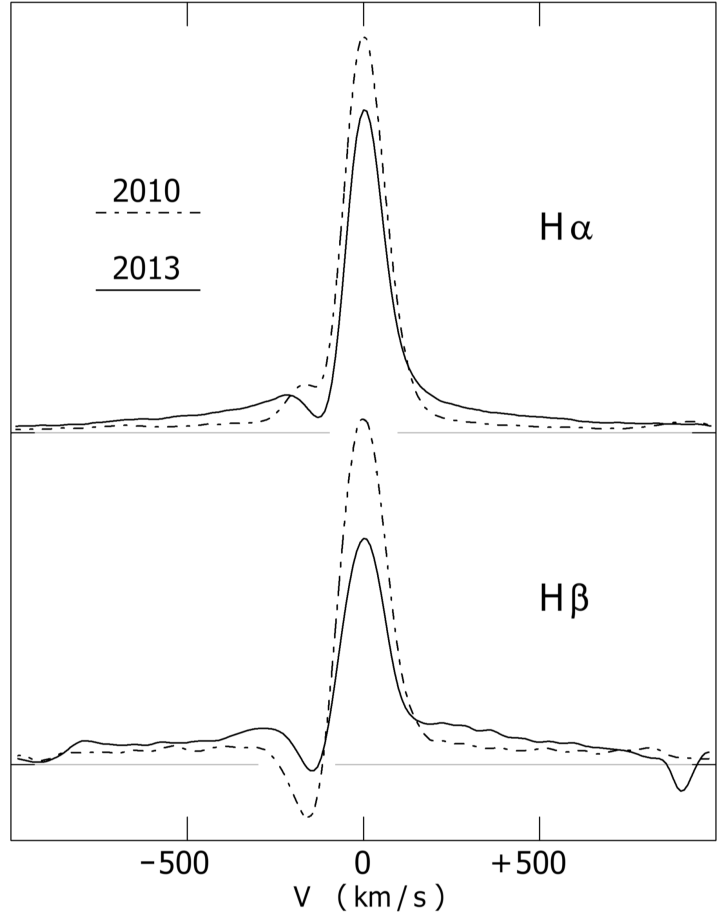}
\caption{The H$\alpha$ and H$\beta$ emission line profiles from 2010 and 2013. Each profile has been normalized so the total area under the curve is the same in 2013 as in 2010. }
\end{figure}

\end{document}